# *How and Why the Universe Began*

*Roger Ellman*

Abstract

    Before the universe there was nothing, absolute nothing. That is the starting point because it is the only starting point that requires no cause, no explanation nor justification of its existence.

    But, that starting point has two impediments to the universe, or anything, coming into existence from it. First is the problem of change from nothing to something without, at least initially, an infinite rate of change, which is impossible. Second is the problem of change from nothing to something without violating conservation, which must be maintained.

    Mathematical analysis develops the sole solution to both problems. That is that the beginning had to be of a *±[1 - Cosine(2π·f·t)]* form. That solution indicates the fundamental nature of matter, energy and field.

Roger Ellman, The-Origin Foundation, Inc.
    320 Gemma Circle, Santa Rosa, CA 95404, USA
    RogerEllman@The-Origin.org
    http://www.The-Origin.org

# *How and Why the Universe Began*

## *Roger Ellman*

We are confronted with an apparently insuperable problem. Before the universe there was nothing, absolute nothing. That is the starting point because it is the only starting point that requires no cause, no explanation nor justification for its existence. But, that starting point has two impediments to the universe, or anything, coming into existence from it. First is the problem of change from nothing to something without, at least initially, an infinite rate of change, which is impossible. Second is the problem of change from nothing to something without violating conservation, which must be maintained.

The analysis would appear to end at that point, end with the declaration that obviously there cannot be a universe and there is no universe. Except, of course, that we and the universe we inhabit clearly exist at least enough for us to investigate it. Therefore, a solution to the insuperable problem exists. That solution is as follows.

## *1 - THE PROBLEM OF INFINITE RATE OF CHANGE*

To avoid a material infinity the rate of change at the moment of the change must have been finite. Rather than an instantaneous jump from nothing to something, no matter how small or "negligible" that something might have been, there had to be a gradual transition at a finite rate of change. Further, the rate of change of that rate of change, the change's second derivative, at that moment had to have been finite, and so on *ad infinitum* for all of the further derivatives.

That requirement means that the form of the change had to have been either a natural exponential or some form of sinusoid. That develops as follows, in which the sought form of the change will be the function $U(t)$ [the "$U$" for universe, of course].

To illustrate the problem consider the following function

(1) $\quad U(t) = 0 \quad\quad t < 0$
$\quad\quad\; U(t) = t^2 \quad\; t = 0 \text{ and } t > 0$

as a theoretical candidate for $U(t)$ at the beginning of the universe, which function is graphically depicted at the right.

Its first derivative, also graphically depicted to the right, is

(2) $\quad \dfrac{dU(t)}{dt} = 0 \quad\quad t < 0$

$\quad\quad\;\; \dfrac{dU(t)}{dt} = 2 \cdot t \quad\; t > 0$

and is unstated for $t=0$ because $dU(t)/dt$ is not smooth there even though $U(t)$ "looks" smooth there.

Now, the second derivative

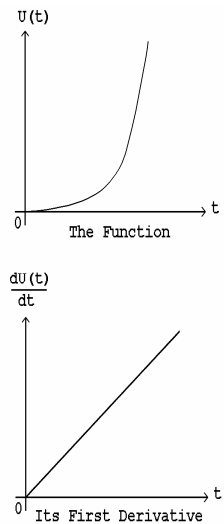

*Figure 1a*

(3) $$\frac{d^2U(t)}{dt^2} = 0 \quad t < 0$$

$$\frac{d^2U(t)}{dt^2} = 2 \quad t > 0$$

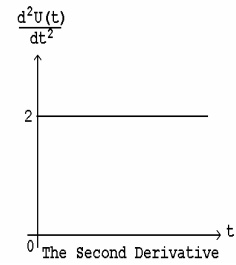

*Figure 1b*

is clearly discontinuous at *t=0*, the instant of the beginning of the universe, where it instantaneously jumps from *0* to *2* as is graphically depicted to the right.

The third derivative, which is the rate of change of the second derivative must be infinite at *t=0* to produce the instantaneous jump from *0* to *2*. Clearly, that cannot have happened in the real universe. It is such a condition which is unacceptable in a candidate function for *U(t)* at the beginning of the universe.

The only way to avoid that condition of an infinite derivative somewhere along the line of successive further derivatives is to have a function with an endless family of finite, non-zero derivatives; that is, some derivatives may be zero at *t=0* but there must always be further non-zero higher derivatives, which requires that the functional form of every derivative must be non-zero.

One can conceive theoretically of the idea of a function for which all derivatives are non-zero and no two are alike (in a general sense analogous to the pattern of digits in an irrational number), but it is not likely that such a function can exist. In any case the more certain and more simple way to achieve all non-zero derivatives is a repeating derivative function, the two simplest examples of which are

(4) $$\frac{dU(t)}{dt} = \pm U(t) \quad \text{[First derivative = the original function]}$$

(5) $$\frac{d^2U(t)}{dt^2} = \pm U(t) \quad \text{[Second derivative = the original function]}$$

### a. *Analysis of Repeating Derivative Functions*

*Case (a): Functions Satisfying Equation (4)*

The function meeting this requirement is the natural exponential, $\varepsilon^t$.

(6) $$\varepsilon^t = 1 + t + \frac{t^2}{2!} + \frac{t^3}{3!} + \ldots$$

Taking the first derivative

(7) $$\frac{d[\varepsilon^t]}{dt} = 0 + 1 + \frac{2t}{2!} + \frac{3t^2}{3!} + \ldots$$

$$= 1 + t + \frac{t^2}{2!} + \frac{t^3}{3!} + \ldots = \varepsilon^t$$

so that the original function results as is required by equation *(4)*.

That is the prime case of a function that satisfies the requirement of all derivatives existing in functional form. In general those of this case are as equation *(8)*.

(8)  $U(t) = A \cdot \varepsilon^t$

The function $\varepsilon^t$ is not suitable for $U(t)$ at the beginning of the universe, however, because its value at $t=0$ is not zero. In fact it is zero only at $t = -\infty$. A function that might seem usable, however, would be

(9)  $U(t) = 0 \qquad t < 0 \text{ and } t = 0$

$U(t) = \varepsilon^t - 1 \qquad t > 0$

$\qquad\qquad = t + \dfrac{t^2}{2!} + \dfrac{t^3}{3!} + \cdots$

which does have zero value at $t=0$ and otherwise meets the derivatives requirement sufficiently.

### Cases (b) – (e): Functions Satisfying Equation (5)

Turning to functions that meet the requirement that the second derivative equal the original function per equation *(5)* there are four such functions.

(10) Case (b):  $U(t) = 1 + \dfrac{t^2}{2!} + \dfrac{t^4}{4!} + \cdots$

(11) Case (c):  $U(t) = 1 - \dfrac{t^2}{2!} + \dfrac{t^4}{4!} + \cdots$

(12) Case (d):  $U(t) = t + \dfrac{t^3}{3!} + \dfrac{t^5}{5!} + \cdots$

(13) Case (e):  $U(t) = t - \dfrac{t^3}{3!} + \dfrac{t^5}{5!} + \cdots$

These five candidate functions can be described and summarized as their exponential equivalents as in Figure 2, below.

| Case | Function | Name of Function | Candidate U(t) |
|---|---|---|---|
| (a) | $\varepsilon^t$ | Natural exponential | $\varepsilon^t - 1$ |
| (b) | $\dfrac{\varepsilon^t + \varepsilon^{-t}}{2}$ | Hyperbolic cosine | $\mathrm{Cosh}(t) - 1$ |
| (c) | $\dfrac{\varepsilon^{i \cdot t} + \varepsilon^{-i \cdot t}}{2i}$ | Cosine | $\mathrm{Cos}(t) - 1$ |
| (d) | $\dfrac{\varepsilon^t - \varepsilon^{-t}}{2}$ | Hyperbolic sine | $\mathrm{Sinh}(t)$ |
| (e) | $\dfrac{\varepsilon^{i \cdot t} - \varepsilon^{-i \cdot t}}{2i}$ | Sine | $\mathrm{Sin}(t)$ |

*Figure 2*

The relationships in the table can be verified by substitution using the formula for $\varepsilon^t$ as given in equation *(6)*, above. Cases *(b)* and *(c)* have the same problem that case *(a)* had, that the value of *U(t)* is not zero at *t=0*. Just as with case *(a)*, they would appear to become satisfactory if a constant, *1*, is subtracted from each of them.

These candidates all satisfactorily meet the requirement for a continuous family of derivatives so that the kind of unacceptable problem as encountered in the example of *U(t)=t²* at the beginning of this paper is avoided. That is, all derivatives are finite. But, there are other requirements that the successful candidate function must meet.

### b. *Using the Remaining Criteria to Select U(t)*

Two other criteria must be met by the successful candidate function or functions:

- the function must not be open-ended, that is it cannot ever have an infinite amplitude, and

- the function must smoothly match the *U(t)=0* condition at *t=0*.

The first criterion eliminates cases *(a)*, *(b)* and *(d)* each of which goes to an infinite value of *U(t)*. To satisfy the second criterion the tangent to *U(t)* at *t=0* must be identical to the tangent to the function for *t < 0*, which is the horizontal *t-axis*. The condition is satisfied if the first derivative of *U(t)* equals *zero* at *t=0*. Only cases *(b)* and *(c)* meet that requirement.

Therefore, the resulting form of *U(t)*, the only acceptable form, the only one that meets all of the requirements, is case *(c)*,

*(14)*   U(t) = [Cos(t) - 1]              t > 0 and t = 0

    U(t) = 0                              t < 0.

which is identical in form to the more usual and convenient equation *(15)*

*(15)*   U(t) = U₀·[1 - Cos(2π·f·t)]

in which an amplitude parameter, $U_0$, and a frequency parameter, *f*, have been added.

That the only possible form for the manner in which the universe began is a sinusoidal oscillatory form would seem to be very appropriate. Oscillations, waves, are ubiquitous in our universe from oceans, violin strings and pendulums to sound, light and electron orbits. Most likely that statement can also be validly inverted: Oscillations and waves are ubiquitous in our universe because the universe began from an initial such oscillatory form.

## 2. *T*HE *P*ROBLEM OF *M*AINTAINING *C*ONSERVATION

At this point, that is the universe having started from absolute nothing as an oscillation having the form of equation *(15)*, the maintaining of conservation, the avoiding of getting something from nothing, clearly could only happen in one manner:

There simultaneously had to have arisen an identical-in-form but opposite-in-amplitude oscillation so that the pair balanced out to the original net nothing, as in equation *(16)*.

*(16)*   U(t) = ± U₀·[1 - Cos(2π·f·t)]

There is no other way that violating the assured principle of conservation could have been avoided. The universe exists. It had to come into being from a prior nothing.

That had to happen while avoiding an infinity of amplitude and an infinity of rate of change. Conservation had to be maintained. Ergo equation *(16)*.

It is not proposed that the "+" aspect of equation *(16)* relates to matter and that the "−" aspect relates to anti-matter. Rather, as is developed further below, our universe's matter is a composite of both as is its anti-matter.

Of course, the beginning so far presented raises many questions, among them are the following.

- Why did that, or any, beginning of the universe happen ?

- What was oscillating, that is what was the substance that the developed form applied to ?

- Why did the effects of equation *(16)* not promptly cancel out to a resumption of on-going absolute nothing ?

- How did that beginning evolve into the universe that we know today, how does it relate to today's universe ?

### 3. *THE PROBLEMS: WHY THAT OSCILLATION BEGAN AND WHAT IT WAS*

#### a. *Why That Beginning happened*

A duration is the period of time that a particular state or set of conditions persists. The duration is terminated by a change, which change also initiates a new duration. In the universe change is ubiquitous. It is the constant and continuous stream of change that makes durations mensurable. Before the beginning of the universe a duration was in process even though it was not mensurable. The beginning of the universe was the first change ever and it terminated the original primal duration of absolute nothing.

The probability of the happening of such an event as the universe beginning in the manner described above was / is extremely small. But the event was / is not impossible. Furthermore, in the absence of that event occurring there was an extremely large duration of opportunity in which that extremely small probability could operate. In the absence of the beginning the original duration would have been infinite and that infinite opportunity operated on by minute, but non-zero, probability results in absolute certainty. The beginning of the universe could not avoid eventually happening.

#### b. *What That Beginning Oscillation Was*

The starting point is the assumption that, when the primal nothing changed as a probabilistically inevitable interruption of what would otherwise have been an infinite duration of the primal nothing, the simplest or minimum conservation-maintaining interruption that could occur is what occurred. There are two reasons for this. Occam's Razor, calls for the simplest hypothesis as the most likely. More importantly, or perhaps the same thing, if an essentially spontaneous and extremely low probability event is to occur solely as an interruption of the duration of an otherwise absolute nothing, then very little interrupting event is needed; the barest minimum of something is sufficient to interrupt, to be a change in absolute nothing. There is no call, no reason for anything more. So, while the interruption could have been otherwise, it was probably as simple and minimum as possible.

Size or amount of time are of no meaning here because there is nothing to which they can be compared or by which they can be measured. Whatever amount of change occurred is what occurred. Whatever time it took, or went on for, whatever its oscillatory frequency was, is what happened. Twice as much or half as much have no meaning.

Three conclusions about the initial oscillatory $U_0 \cdot [1 - Cos(2\pi \cdot f \cdot t)]$ form can now be reasonably obtained:

- clearly the universe of today must be an on-going evolved consequence of its beginning, of the initial oscillatory form;

- the frequency, $f$, of the sinusoidal oscillation was, and is, very large; and

- the nature of the change is one of concentration or density of the something that is oscillating.

The frequency would have to be either very large or very small -- high enough so that it is not detected or noticed by us in every day life or so low that it appears to us as no change at all in our experience.

It has already been noted that the fact that the only possible form for the manner in which the universe began is a sinusoidal oscillatory form is very appropriate because oscillations, waves, are ubiquitous in our universe from oceans, violin strings and pendulums to sound, light and electron orbits. And it has been noted that that statement can be validly inverted: oscillations and waves are ubiquitous in our universe because the universe began from an initial such oscillatory form.

If the frequency of the initial oscillation were so small that it appears to us as no change at all it would completely eliminate oscillations playing any significant part in the behavior of the universe as we know it. Therefore, the frequency must have been very large, so rapid compared to our perception that we do not notice the oscillation at all.

The change can hardly be one of gross size if it is going on right now at high frequency as has just been concluded. One can conceive of the fundamental "substance", the "something" of the universe flashing into and out of existence from a zero to a maximum density or concentration in an oscillatory fashion at a rate so high that we neither detect nor notice it at all. But, it is not possible to entertain a concept of reality flashing from zero to full size, a size that includes ourselves and our environment, in such a fashion.

Actually, the reality that we know is not "flashing into and out of existence ...." Our reality is more the oscillation itself than what is oscillating and the continuing oscillation is our steady, constant reality.

Thus the interruption that gave us our universe was the starting of an oscillation, present to us at a very high frequency and of $U_0 \cdot [1 - Cos(2\pi \cdot f \cdot t)]$ form, of the *density*, as the variation will be hereafter referred to, of the *medium*, as what it is that is oscillating will be hereafter referred to. Now, two other questions must be addressed:

- What about the identical-in-form but opposite-in-amplitude oscillation that maintains conservation and

- What is the medium, what is that which is oscillating ?

All of the discussion so far must apply to the "negative" oscillation, $-U(t)$, exactly as to the "positive" oscillation $+U(t)$ because the exact same reasoning as for $+U(t)$ applies to $-U(t)$ and, after all, they are not distinguishable in the discussion. The terms "$+$" and "$-$" are merely terms of convenience for two equal form opposite magnitude unknown things. We probably tend to think of our universe as the "$+$", but that is meaningless and irrelevant. There can be no objective designation of $+U(t)$ and $-U(t)$, no way to identify one versus the other. Both had to appear and our universe cannot avoid being the evolved result of both.

The question could arise as to whether  $+U(t)$  and  $-U(t)$  are co-located or separate.  The answer is that they must be co-located.  Their function relative to each other is to maintain overall conservation from the beginning.  That conservation must be maintained locally and generally, which requires that they occupy the same space.  They initially are identical except for their  $+/-$  oppositeness and therefore each must obey the same laws thereafter.  Those laws practice conservation, consequently conservation will be maintained if the beginning conserved, which it did.

Since  $+U(t)$  and  $-U(t)$  are co-located the universe that we know and exist in is the combined integrated result of both  $+U(t)$  and  $-U(t)$.  The "+" and "−" electric charges of our universe [in both matter as for example in protons and electrons and in anti-matter as for example in negaprotons and positrons] must derive from that aspect of the beginning.  (It is interesting to observe, also, that our universe being the integrated result of an initial beginning and its opposite relates to (presumably is the underlying cause of) the dialectical nature of reality, the ying and yang of oriental philosophy.)

The question of what the medium is can only be answered in terms of its characteristics, what it does and how.  It will be fruitless to attempt to use human terms (gas, jelly, field, aether, or whatever) to describe that which so far underlies the things our vocabulary was developed to describe.  We might as well call it a chocolate sauce or a raspberry mousse.  The characteristics of the medium are its definition.  The medium is:

- a continuous entity, not a mass of "particles" nor anything having parts;

- simple and uniform throughout (except, of course, for the density variations, the oscillation); and

- of minimum "tangibility" or "substantiality".

## 4. THE PROBLEMS:

- *WHY DID THE EFFECTS OF EQUATION (16) NOT CANCEL AND ON-GOING ABSOLUTE NOTHING RESUME ?*

- *HOW DID THAT BEGINNING PRODUCE THE "BIG BANG" AND EVOLVE INTO OUR UNIVERSE OF TODAY ?*

### a. Development of the Unified Field

To resolve those questions it is necessary to develop some aspects of our present material reality.  The following "thought experiments" develop the concepts.

Electric Field

- Nothing can travel faster than the speed of light,  $c$.  Given two static electric charges separated and with the usual Coulomb force between them, if one of the charges is moved the change can produce no effect on the other charge until a time equal to the distance between them divided by  $c$  has elapsed.

- For that time delay to happen there must be something flowing from the one charge to the other at speed  $c$  and the charge must be the source of that flow.

    The Coulomb effect is radially outward from the charge, therefore every charge must be propagating such a flow radially outward in all directions from itself, which flow must be the "electric field".

Motion of Charge and "At Rest"

- Comparing two such charges, one moving at constant velocity relative to the other, at least one of the charges is moving with some velocity,  $v$.

- The flow (of "field") outward from that charge must always travel at $c$. Forward it would go at $[c + v]$ if propagated at $c$ from the source charge already moving that way at $v$. Therefore, it must be sent forward from the charge at $[c - v]$ so that it will travel at $c$ when the $v$ of its source charge is added.

*(17)* $\text{Flow}_{\text{fwd}} = [c - v]$

- Analogously, rearward it would go at speed $[c - v]$ if propagated at $c$ from the source charge already moving the opposite way at $v$. Therefore, it must be emitted rearward from the charge at $[c + v]$ so that it will travel at speed $c$ when the $v$ of the source charge in the opposite direction is subtracted.

*(18)* $\text{Flow}_{\text{rwd}} = [c + v]$

- But, that rearward – forward differential means that the direction and speed of motion can be determined by looking at the propagation pattern of the flow as propagated by the charge.

    And, if the pattern were the same in all directions then the charge would be truly "at rest", which means that there is an absolute "at rest" frame of reference.

## Unification of Fields

- Except for the kind of field, all of the preceding applies in the same way and with the same conclusions for magnetic field and gravitational field as for electric field.

- Therefore, either a particle that exhibits all three such fields, as for example a proton or an electron, is a source of three separate and distinct such flows, one for each field, or there is only a single such flow which produces all three effects: electric, magnetic, and gravitational.

    The only reasonable conclusion is that electric, magnetic, and gravitational field are different effects of the same sole flow from the source particles.

## Sources & Their Decay

- The flow is not inconsequential. Rather, it accounts for the forces, actions and energies of our universe.

- For a particle to emit such a flow the particle must be a source of whatever it is that is emitted outward. The particle must have a supply of it.

- The process of emitting the flow from a particle must deplete the supply resource for the particle's emitting further flow, must use up part of its supply, else we would have something-from-nothing and a violation of conservation.

    It must be concluded that an original supply of that which is flowing came into existence at the beginning of the universe and has since been gradually being depleted at each particle by its on-going outward flow.

## That Which is Flowing

- The flow is a property of contemporary particles. Those particles are evolved successors to the original oscillations with which the universe began. Then, that which is flowing is the same original primal "medium", the substance of the original oscillations at the beginning of the universe.

    Since it is flowing outward from the myriad particles of the universe simultaneously and that flow is interacting with myriad others of those particles without untoward interference, the "medium" must be extremely intangible for all of that to take place, any one particle's flow flowing

largely freely through that of other particles, as intangible as -- well -- "field".

## The Beginning

- Before the universe began there was no universe. Immediately afterward there was the initial supply of medium to be propagated by particles. How can one get from the former to the latter while: (1) not involving an infinite rate of change, and (2) maintaining conservation ?

   The only form that can accommodate the change from nothing to something in a smooth transition without an infinite rate of change is the oscillatory form of equation *(19a)*, below.

*(19a)*  $U_0 \cdot [1 - \cos(2\pi \cdot f \cdot t)]$

   The only way that such an oscillation can have come into existence without violating conservation is for there simultaneously to have come into existence a second oscillation, the negative of equation *(19a)* as in equation *(19b)*.

*(19b)*  $-U_0 \cdot [1 - \cos(2\pi \cdot f \cdot t)]$

   That is, the two simultaneous oscillations must have been such as to yield a net of nothing, the prior starting point, when taken together.

## The Oscillatory Medium Flow ≡ Electric charge and field

- The initial medium supply of each particle, each being a direct "descendant" of the original oscillation at the universe's beginning, must be oscillatory in form per equations *(19)*. Therefore the radially outward flow from each particle is likewise an oscillatory medium flow of the form of equations *(19)*.

   The flow is radially outward from the particle, therefore, the oscillation of the medium supply of each particle is a spherical oscillation. The particle can also be termed a *center-of-oscillation*, which term will also be used here.

- The amplitude, $U_0$, of the *[1-Cosine]* form oscillation is the amplitude of the flow emitted from the source particle, which flow corresponds to the electric field. Thus the oscillation amplitude must be the charge magnitude of the source particle -- the fundamental electric charge, $q$, in the case of the fundamental particles, the electron and the proton. See[2] for more on Coulomb's Law and inertial mass.

   Then, the conservation-maintaining distinction of amplitude $+U_0$ versus amplitude $-U_0$ must be the positive / negative charge distinction.

   The frequency, $f$, of the *[1-Cosine]* form oscillation must then correspond to the energy and mass of the source particle, that is the energy of the oscillation is $E = h \cdot f$ and the mass is $m = E/c^2 = h \cdot f/c^2$.

[- While it does not pertain to the universe's beginning, because the outward medium flow from each particle must deplete the source particle's remaining supply of medium for further propagation, the amplitude magnitude, $U_0$, must exponentially decay. That is, it must be of the form of equation *(20)*, below.]

*(20)*  $|U(t)| = U_0 \cdot \varepsilon^{-t/\tau}$

## Medium Emission and Medium Flow

- When a charge is at rest, medium is emitted by it and flows outward in the same manner in all directions. But, when the charge is in motion at constant velocity, $v$, the flow forward is emitted at speed $[c - v]$ and rearward at $[c + v]$ per above.

- There can be only one frequency, $f$, in the $[1 - Cosine(2\pi \cdot f \cdot t)]$ form oscillation of the emitted flow regardless of whether it is directed forward, rearward or sideward. Therefore, to obtain the slower speed, $[c - v]$, emitted forward the wavelength forward, $\lambda_{fwd}$, must be shorter so that the speed at which the flow is <u>emitted</u>, $= f \cdot \lambda_{fwd}$, will be slower.

- The case is analogous rearward where $\lambda_{rwd}$ is longer in order for the speed, $[c + v]$, to be greater.

> In all directions from the moving charge, including any that are partially sideward plus partially forward or rearward, the speed of emission and the wavelength emitted will be the resultant of the sideward plus forward or rearward components of a ray in that direction.

- The absolute rate of flow outward of the emitted medium must be at speed $c$. Forward that comes about because the forward speed of the charge, $v$, adds to the forward speed at which the medium is emitted, $[c - v]$, resulting in the medium flowing at the speed of the sum, $speed = v + [c - v] = c$.

- That speed increase raises the $[1 - Cosine(2\pi \cdot f \cdot t)]$ form oscillation frequency (per the Doppler Effect). Thus forward medium <u>flow</u> speed is $c = f_{fwd} \cdot \lambda_{fwd}$.

- Analogously rearward the speed of medium <u>flow</u> is at $c = f_{rwd} \cdot \lambda_{rwd}$.

> In all directions from the moving charge, including any that are partially sideward plus partially forward or rearward, the speed of flow will be $c$ and the frequency and wavelength of the flow will be the resultant of the sideward plus forward or rearward components of a ray in that direction.

Magnetic Field

- A charge at rest exhibits the electrostatic effect but not the magnetic effect. That charge has a spherically uniform pattern of $[1 - Cosine(2\pi \cdot f \cdot t)]$ form oscillatory medium emission and flow outward.

- A charge in motion exhibits the magnetic effect in addition to the electrostatic effect. That charge has a pattern of emission and outward flow of medium that is cylindrically symmetrical about the direction of motion but that varies in wavelength and frequency from $f_{fwd} \cdot \lambda_{fwd}$ forward to $f_{rwd} \cdot \lambda_{rwd}$ rearward.

> The electrostatic [Coulomb's Law] effect is due to charge location. The magnetic [Ampere's Law] effect is due to charge motion. Clearly, then, the electrostatic effect is due to the spherically uniform medium flow from the charge and the magnetic effect is due to the change in shape of that medium flow pattern caused by the charge's motion.

Electro-magnetic Field

- There is a continuous emission of medium in $[1 - Cosine(2\pi \cdot f \cdot t)]$ oscillatory form from each charge, which medium flows outward, away, forever. Constant velocity motion of a charge produces a change in the frequency and wavelength of that medium flow.

- Changes in the velocity of the charge cause corresponding further changes in the medium's oscillatory form as successive increments of medium are emitted and flow outward from the charge. Earlier increments so changed propagate on outward away from the charge, forever, at $c$. The stream of outward flowing medium carries a history of the motions of the source charge.

> Propagating electromagnetic field is the carrying of both of those field aspects as an imprint on the otherwise uniform medium flow from the charged particle, an imprint analogous to the modulation of a carrier wave in radio communications.

- Electro-magnetic field is caused by acceleration / deceleration of charge, that is by changes in the charge velocity. Therefore:

> The changing electric and magnetic fields of electro-magnetic field actually <u>are</u> form changes imprinted onto the outgoing medium flow and carried passively with it [analogous to modulation of a carrier wave in radio communications].

> Because all medium flow is spherically outward in all directions from its source charge, changes in it, caused by changes in the source velocity, propagate outward <u>in all directions</u>. Those medium flow changes <u>are the</u> changing electric and changing magnetic fields of <u>electro-magnetic field.</u>

- It is not the speed of light which is the fundamental constant, $c$, light being a mere modulatory imprint on medium flow. It is the speed of medium flow which is the fundamental constant, $c$.

Gravitational Field

- As pointed out earlier above, the frequency, $f$, of the $[1 - Cosine]$ form oscillation corresponds to the mass of the source particle. Therefore the frequency aspect of the radially outward medium flow is the "gravitational field". See [3] for more on gravitation.

### b. *Further Analysis of the Beginning's Initial Oscillations*

The analysis so far has developed the only form that can accommodate the requirements of the beginning of the universe, change from nothing to something in a smooth transition without an infinite rate of change while also not violating conservation, which form is per equation *(16)* which is also equations *(19a)* and *(19b)* and appears graphically as Figure 3, below.

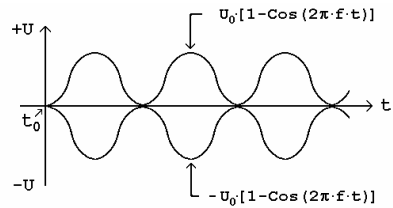

*Figure 3*

Examination of this wave form reveals two problems. One, that it is an immediate mutual annihilation, will be dealt with shortly below. Of concern now is that an infinite rate of change still remains. The envelope of the oscillation has an infinite rate of change at $t=t_0$ as can be seen in Figure 4, below, which displays the envelope. That infinite rate of change is no more acceptable than that of the original problem.

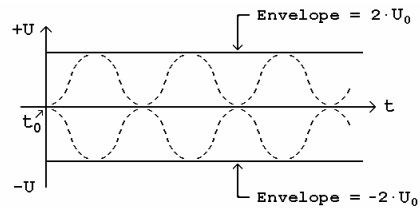

*Figure 4*

Viewed in a mathematical or graphical sense without any consideration of the physical reality represented, the envelope discontinuity at $t = t_0$ is not a difficulty. The only quantity that actually exists and is varying is the overall $U(t)$. The envelope is merely our perception of a characteristic of the wave form. The actual varying quantity, per Figure 3, has no discontinuity at $t = t_0$ for the reasons already presented.

However, looked at in a physical sense the oscillations depicted in Figure 3 are the effects called *energy*, *mass* and *charge* embodied in what we call a "particle" and are something other than nothing. They are a material physical reality that did not exist prior to the beginning of the universe. They can no more leap from zero to a finite non-zero amount than could the original $U(t)$ so leap. That infinite rate of change in the amount of *energy / mass / charge / oscillation* at $t = t_0$ is no more acceptable than was the infinite rate of change encountered in the original analysis of the probable beginning and it must be corrected by the same kind of reasoning as then pursued: the envelope, also, had to originate as a *[1 - Cosine]* form of oscillation, the only form that avoids an infinite rate of change and matches the requirements of the situation.

That original envelope oscillation was at a lesser frequency than the original wave by the definition of a wave form envelope. If it were at a greater frequency then the roles (envelope and wave) would be reversed. If it were at the same frequency it would not act as an envelope and the infinity problem would remain. If we designate the envelope frequency as $f_{env}$ and the frequency of the wave oscillation within the envelope as $f_{wve}$ then the envelope would be of the following form.

*(21)*  $U_{env} = [1 - \text{Cos}(2\pi \cdot f_{env} \cdot t)]$

The wave is, as before, of the form

*(22)*  $U_{wve} = \pm U_0 \cdot [1 - \text{Cos}(2\pi \cdot f_{wve} \cdot t)]$

and the envelope-modulated wave is per equation *(23)* and Figure 5, below.

*(23)*  $U(t) = [U_{env}] \cdot [U_{wve}]$

$\qquad = \pm U_0 \cdot [1 - \text{Cos}(2\pi \cdot f_{env} \cdot t)] \cdot [1 - \text{Cos}(2\pi \cdot f_{wve} \cdot t)]$

The "±" in the expression accounts for the oscillation being of both *+U(t)* and *-U(t)*, of course, so that conservation is maintained.

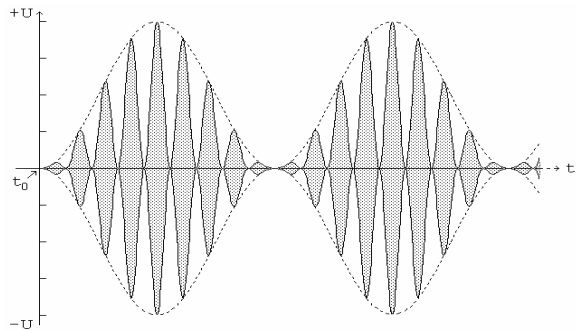

*Figure 5*

There is no contention here that *+U(t)* represents "positive" *energy / mass* and *-U(t)* negative. Neither the mass effect nor the energy effect is a "real" reality. The only reality is the oscillations; all else is our perception of the effects that are produced by the centers and their waves. Among those effects are what we have chosen to refer to as *mass* and *energy*. The only reality, the oscillation, consists of two equal and opposite oscillations that mutually maintain conservation.

There is no such thing as negative mass or energy. (Negative energy amounts are spoken of in physics discussions but they are not absolutely negative, only negative relative to some other defined energy. For example, the energy of an atom's orbital electrons is negative relative to the energy that they would have if they were free of the atom.)

Photons and electromagnetic waves carry energy and it is always "positive". That is our perception of the effects that they produce. Actually, so far as the underlying outward propagating medium wave on which the photon E-M wave is a modulating imprint is concerned, a photon deriving from a particle / center-of-oscillation in $-U(t)$ is $180°$ out of phase with one deriving from a $+U(t)$ particle / center-of-oscillation, and reflects the same conservation as the original $+U(t)$ and $-U(t)$ oscillations. Likewise, electro-magnetic radiation from a positive particle's motion is $180°$ out of phase with E-M radiation from a correspondingly moving negative particle.

However, the form of $U(t)$ of equation *(23)* and Figure 5 still does not resolve the problem of an infinite rate of change at $t=t_0$. The *[1 - Cosine]* envelope is itself an oscillation that begins at $t_0$ with a sudden step from zero to its full amplitude. Figure 5 shows the first two cycles of that envelope oscillation, which if only the envelope is considered, is a simple oscillation at the envelope frequency, even though visually, in the figure, it is only the trace of the peaks of the overall complex oscillation. It is *energy / mass / oscillation* that begins suddenly in its full amount at $t_0$ just as, in Figure 3, the oscillation of equation *(21)* begins at $t_0$.

Therefore, it is necessary to introduce yet another envelope of *[1 - Cosine]* form to prevent the infinite rate of change at $t_0$ in the prior envelope. That correction will in turn require still another such correction and so *ad infinitum*. Apparently at this point an infinite string of envelopes thus results as a necessity of the situation. The resulting $U(t)$ would then be

*(24)*
$$U(t) = \pm U_0 \cdot \prod_{i=1}^{i=\infty} [1- \cos(2\pi \cdot f_{env_i} \cdot t)] \times \cdots$$

$$\cdots \times [1- \cos(2\pi \cdot f_{wve} \cdot t)]$$

where the $\prod$ symbol (a large π, Greek "p")
means the product of the indicated factors the "Cosmic Egg" except that several other considerations further modify the situation.

(1) While the foregoing reasoning is sequential - from an original wave to an original wave and its envelope, then to a second envelope, and then another *ad infinitum* - the event was instantaneous. Analogous to the manner in which a cosine function, which has an infinite set of derivatives (which are the means by which it avoids an infinite rate of change), springs "full blown" into existence rather than occurring as the function followed by the first derivative, then the second, and so forth; so the overall original oscillation, $\pm U(t)$ with its infinite set of envelopes also had to spring "full blown" into existence, not appearing first with one, then a second, and so forth, envelopes.

The unending series of successive derivatives of a cosine results nevertheless in a limited or closed form, the cosine. It can be represented by an infinite series of terms which, because each successive term is sufficiently less than the prior term, has a definite sum, the cosine (i.e. the series is convergent). But, it would appear that the infinite series of envelopes of $U(t)$, while theoretically necessary, cannot exist in a real physical situation. There must have been some kind of convergence to a definite, limited sum or form. Furthermore each additional envelope corresponds to an additional increment of *energy / mass* and there cannot be an infinite amount of that. Something had to set a finite limit on the number of envelopes.

(2) Only the "outer" or "last" envelope being of the `[1 - Cosine]` form is necessary to control the difficulty of an infinite rate of change at $t_0$. All of the "inner" envelopes, and the wave itself, being simple cosines rather than `[1 - Cosine]` forms is far more simple. It would appear to have been the more likely actual case. That assumption is not essential to the following development, but it does make its presentation far simpler and more practical.

[It turns out that whether the "`[1 - `" part of the `[1 - Cosine]` form of the envelopes, is present or not (other than that of the "last" one) the net effect on the form of the "Cosmic Egg" oscillation is the same because the number of envelopes is so extremely large. That is demonstrated in Figures 6 on the following page, which shows the convergence of the two different wave forms, that of `[1 - Cos(x)]`$^n$ versus that of `[1 - Cos(x)]·Cos`$^{n-1}$`(x)`, into the same wave form for moderately large $n$, $n=100$, which is still far smaller than the extremely large actual number.]

(3) Each additional envelope factor in equation *(24)* results in a higher frequency content in the overall expression. That is, as each envelope is added the expansion of the expression as the product of multiple cosines into a sum of individual frequency cosine terms becomes longer and acquires higher frequency terms. But, the original oscillation could not have had an actual component at infinite frequency. [Table 8, further below, shows that as a cosine is raised to successively higher integer exponents the highest frequency component in the expansion also increases correspondingly.]

Considering sound waves propagating in a gas as an analogy, there is an upper limit to the frequency of sound that can be propagated. The limit is set in two ways. The wave length of the sound waves decreases as the frequency increases. When the wave length becomes reduced to on the order of the size of the individual particles of the gas it cannot further reduce because the particles cannot subdivide.

Likewise, as the frequency increases the oscillatory motion of the gas particles must become more rapid. But the mass of the particles makes more rapid motion ever more difficult and the motion is ultimately limited by the speed of light. Thus the nature of the medium in which sound waves propagate inherently sets a limit on the propagation of sound waves in that medium.

It is reasonable that there be some aspect of the medium, which, as we know, already limits the speed of medium wave propagation to the speed of light, which aspect sets a limit on the highest frequency / lowest wave length waves that can propagate as medium. That must be the case if for no other reason than to again avoid an infinity and as a result the series of envelopes, of factors in equation *(24)*, was limited to some finite but quite large amount. The real universe original $U(t)$ had an enormous set of envelopes but not an infinite set; they were cut off at some point. (Further analysis of this cutting off is presented later below.)

This reasoning yields a revised $U(t)$, the form of the original oscillation, the "Cosmic Egg", as equation *(25)*, below. There $N_0$ is the number of envelopes. The "`[1 - `" parts, have been eliminated from all but the "$\infty^{th}$" envelope (the "most infinite", the "last" or "outer", envelope), and that envelope does not appear in the expression because the envelopes effectively cut off long before that point.

The resulting form of $U(t)$, the "Cosmic Egg", is as follows.

*(25)* $\quad U(t) = \pm U_0 \cdot [\text{Cos}^{N_0}(2\pi \cdot f_{env} \cdot t)] \cdot [\text{Cos}(2\pi \cdot f_{wve} \cdot t)]$

`        [N`$_0$` is the number of envelopes until they cut off.]`

*Figure 6*
*Comparison of U(t) "Cosmic Egg" Wave Forms (Amplitude Normalized)*

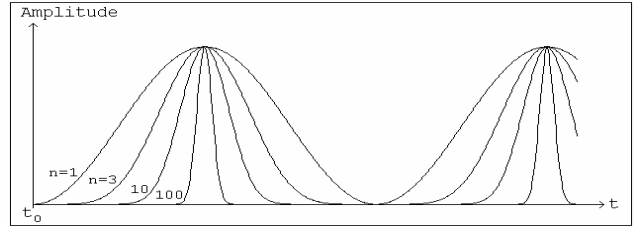

*a. $[1 - Cos(x)]^n$ For n = 1, 3, 10, 100*

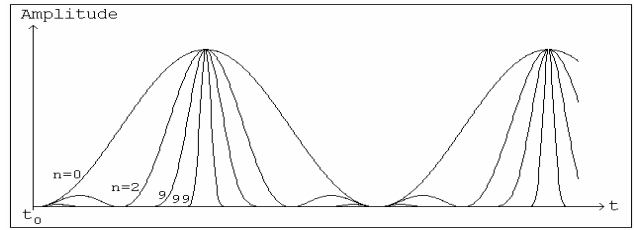

*b. $[1 - Cos(x)] \cdot [Cos(x)]^n$ For n = 0, 2, 9, 99*

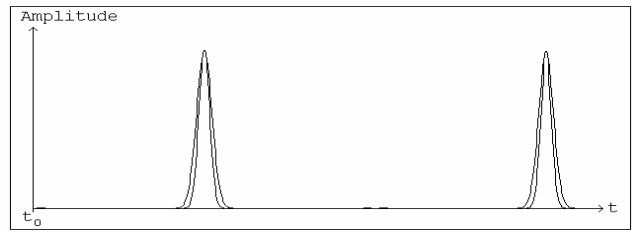

*c. $[1 - Cos(x)]^{100}$ versus $[1 - Cos(x)] \cdot [Cos(x)]^{99}$*

For very large $n$, that is very large $N_0$ of equation *(25)*, the converging of the wave form into a single narrow peak proceeds to a momentary "spike" per cycle. ($N_0$ is found further below to be about $10^{85}$.) Figure 7, below, shows the appearance of the wave form for extremely large $n$, that is for $n = N_0$ it shows what the wave form of the original "Cosmic Egg", the start of our universe, "looked like".

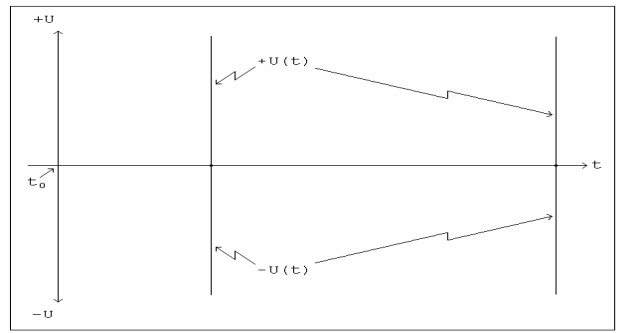

*Figure 7*
*The U(t) "Cosmic Egg" Wave Form of Equation (25)*

As noted above, our today's particles being direct "descendants" of the original oscillation, their initial medium supply must be oscillatory in form per equations *(19)*, the underlying basic form of equation (25) the "Cosmic Egg". Then the "Cosmic Egg" was an immense particle, a super neutron [because electrically neutral], the most complex particle possible. Its atomic mass number was approximately $N_0 = 10^{85}$ as compared to those of the heaviest atomic nuclei which are in the range of *2 - 3 hundred*.

That "Cosmic Egg" was so beyond the level of instability of the heaviest atoms known, in terms of atomic mass number a relative level of instability of $10^{85}$ compared to $10^2$, that it exploded instantly in a massive radiation of particles and energy – the "Big Bang". In terms of Figure 7, that explosive decay happened sufficiently before the wave form reached the first "spike", happened when the progress of mutual annihilation had only partially progressed, and therefore mutually annihilated only a portion of the total wave form leaving the particles that constitute today's universe.

And that is how the only possible beginning gave rise to the "Big Bang". And that is why it did not completely mutually annihilate to no net universe.

### *c* *The Finite Limitation of the "Cosmic Egg" Envelopes*

By "finite limitation" is meant that in the vicinity of the cut-off number of envelopes, $N_0$, the amplitude of each of the further successive envelopes being imposed on the original $U(t)$ was progressively less than its predecessor and the rate of that amplitude decrease increased sharply with further envelopes -- there was a sharp cut off of amplitude and the progressive amplitude of further envelopes approached zero, became infinitesimal. Two effects jointly contributed to there being such a sharp cut-off of the otherwise infinite number of original "Cosmic Egg" envelopes. The first, and most important was a <u>bandwidth effect</u>. The second results from the <u>mathematics of $U(t)$ effect</u>.

The <u>bandwidth effect</u> is exactly analogous to the bandwidth limitation found in electronic devices. An example is sound systems for human use. Such systems are unable to process signals of all frequencies because unavoidable capacitances and inductances in the devices set limits. Such devices always have bandwidths, limited ranges of frequencies that they can successfully process, which are determined by their components and design. In the case of the "Cosmic Egg" a similar bandwidth type of limitation operated.

Every oscillation in nature exhibits, and the theory of oscillations requires, that the oscillation consist of two aspects storing and exchanging the energy of the oscillation back and forth by means of a "flow". With one aspect varying in oscillatory fashion then when that aspect decreases there must be some "place" for its energy to go, a place in which it is stored until it reappears in that aspect when it increases again. It cannot completely disappear or be lost because the oscillation would die. That "place" is the oscillation's second aspect and it obviously must vary in a manner related to the first aspect's variation, but with its energy storage in opposite phase.

A pendulum, for example, oscillates by the motion (flow) of its swinging mass between peak height in the gravitational field (potential energy) at each end of the swing and peak speed of motion (kinetic energy) at the mid-point between the ends of the swing. Then the original oscillation at the start of the universe must have so been: medium being stored or "expressed" in two alternative forms, each oscillating and storing energy in opposite phase to the other.

An example of the general form is an electric circuit having inductance and capacitance exchanging oscillatory energy via flow of electric current as follows.

(26) $$L \cdot \frac{d^2 i}{dt^2} + R \cdot \frac{di}{dt} + \frac{1}{C} \cdot i = 0 \qquad [L \text{ is inductance, } C \text{ capacitance, variable } i \text{ the electric current and } R \text{ is electric resistance.}]$$

The first and third terms of equation *(26)* are the two energy storage processes alternating in opposite phase to each other. The second term treats the loss of energy to the system, for example dissipated or, for the universe's medium, energy lost to the oscillation by the outward flow of medium. The derivatives of the flow, $i$, are functions of the frequency content of that flow.

The definition of bandwidth is that point, for various different frequencies, where the energy [proportional to the square of the flow] becomes reduced to half the peak value and that point is a function of the parameters. The upper bandwidth point occurs at the value of frequency, $f$, for which $R = 2\pi \cdot f \cdot L$. [The lower, not of interest here and very minute for $U(t)$, occurs at $R = 1/2\pi \cdot f \cdot C$.]

The upper bandwidth point for $U(t)$ [at its equivalent to the electrical example's $f = R/2\pi \cdot L$] produced the cutting off of the otherwise infinite series of envelopes in the universe's origination. See [1] for calculation of the "Cosmic Egg" bandwidth, which calculation is too lengthy and involved to be included here, but which correlates well with the value of $N_0$ determined shortly further below.

The second effect, the <u>mathematics of $U(t)$</u> sharpened the cut-off; it made the falling off of amplitude much more drastic once it started. The key to that behavior is to be found in Table 8, below, the expansion of the $Cos^n(x)$ function.

$$Cos^1(x) = 1 \cdot [\ + Cos(x)\ ]$$

$$Cos^2(x) = \frac{1}{2} \cdot [\ 1\ \ \ \ \ \ \ \ \ \ +\ Cos(2x)\ ]$$

$$Cos^3(x) = \frac{1}{4} \cdot [\ + 3Cos(x)\ \ \ \ \ \ \ \ \ \ + Cos(3x)\ ]$$

$$Cos^4(x) = \frac{1}{8} \cdot [\ 3\ \ \ \ \ \ \ + 4Cos(2x)\ \ \ \ \ \ \ \ + Cos(4x)\ ]$$

$$Cos^n(x) = \frac{1}{2^{n-1}} \cdot [\ \cdots\ \ \ \ \ \ \ \ \ \ \ \ \ \ \ \ \ \ \ + Cos(n \cdot x)\ ]$$

*Table 8*
*Expansion of Exponentiated Cosines*

The table makes clear that in the expansion of $Cos^n(x)$ the highest frequency multiple of the fundamental frequency, $x$, is $n$ times that frequency. The "Cosmic Egg" expression equation *(26)* contains the factor

*(27)* $\ \ Cos^{N_0}[2\pi \cdot f_{env} \cdot t]$

and that factor creates the set of envelopes of the original oscillation and means that the highest frequency in the original $U(t)$ was $N_0 \cdot 2\pi \cdot f_{env}$.

The above table illustrates important aspects of the equation *(27)* expansion. The sum of the coefficients of the terms in the expansion always equals the divisor in front of the expression. As a result the expansion has the same overall amplitude, $U_0$, as the unexpanded function (obviously a mathematical necessity). The table also illustrates that the relative amplitude of the increasingly higher frequency terms is increasingly smaller, an effect adding to the bandwidth cut off effect.

Analysis of the coefficient patterns in the terms of the $Cos^n(x)$ expansion discloses a pattern related to the binomial expansion as demonstrated in Table 9, below.

*(a) Binomial Expansion Coefficients $[a + b]^n$*

```
n                                    Coefficients
0                                          1
1                                    1  |  1
2                                 1     2  |  1
3                             1     3     3  |  1
4                         1     4     6  |  4     1
5                     1     5    10    10  |  5     1
6                 1     6    15    20  | 15     6     1
7    1            7    21    35    35  | 21     7     1
:
```

*(b) $Cos^n(x)$ Expansion Coefficients*

```
n                                    Coefficients
         Times Cos(*), * =  0x    1x    2x    3x    4x    5x    6x    7x
0                           1
1                           -     1
2                           1     -     1
3                           -     3     -     1
4                           3     -     4     -     1
5                           -    10     -     5     -     1
6                          10     -    15     -     6     -     1
7                           -    35     -    21     -     7     -     1
:
```

*Table 9*

Clearly, with the exception of the constant term (where, in the table, $* = 0x$) the other terms of the expansion of $Cos^n(x)$ have the same coefficients as the corresponding terms of the binomial expansion. (Of course they must then be multiplied by $1/2^{n-1}$ per Table 8.) The formula for the binomial expansion can thus be used to obtain the coefficients for any value of $n$ in the expansion of $Cos^n(x)$.

In the above table, $N_0 = 10^{85}$ is the $n$ of the formula. It is not practicable to calculate all of the coefficients of the cosine expansion of the envelopes for $10^{85}$ envelopes. On the other hand, it is not unreasonable to calculate the $85$ cases corresponding to the frequency multiples of the expansion: $10^1, 10^2, 10^3, \cdots 10^{85}$.

Figure 10 on the next page is a plot of the relative magnitude of the successive coefficients of the various frequency multiples $(1 \cdot x, 3 \cdot x, \cdots 10^{85} \cdot x)$, in the expansion of $Cos^n(x)$ for $n = N_0 = 10^{85}$. The plot indicates a sharp and drastic cut off, an attenuation of the higher frequencies. Figure 10(a) uses a linear horizontal axis and shows the cut-off in detail. Figure 10(b) uses a logarithmic horizontal scale to better present the tremendous range in frequency multiples from $1$ to $10^{85}$. It shows that the cut-off is quite sharp and drastic.

This cut-off is merely the action of the mathematics of $cos^n(x)$. The complete actual cut-off of the "Cosmic Egg" was the product of this cut-off and the bandwidth limitation discussed above. If this effect operated in the case of an electronic sound system then, with increasing sound frequency, at the approach to the cut-off sound would suddenly cease rather than fade away in reducing amplitude as the bandwidth limitation, alone, causes.

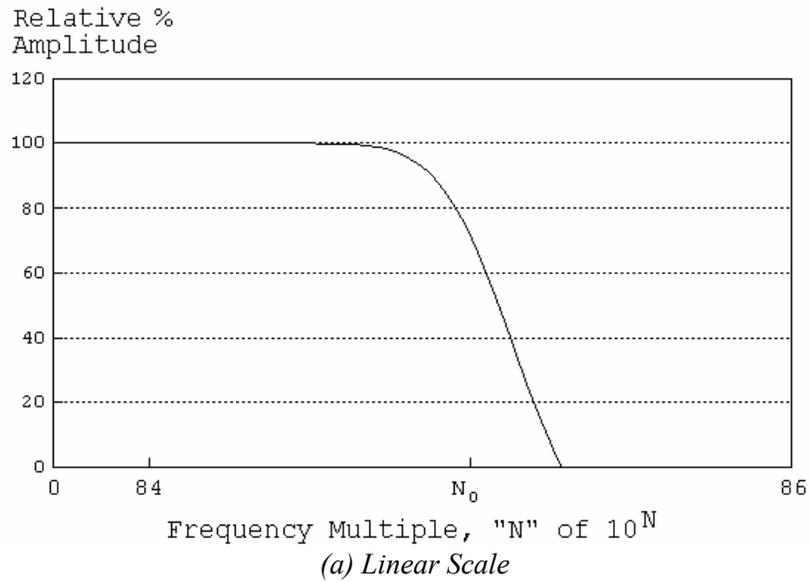

*(a) Linear Scale*

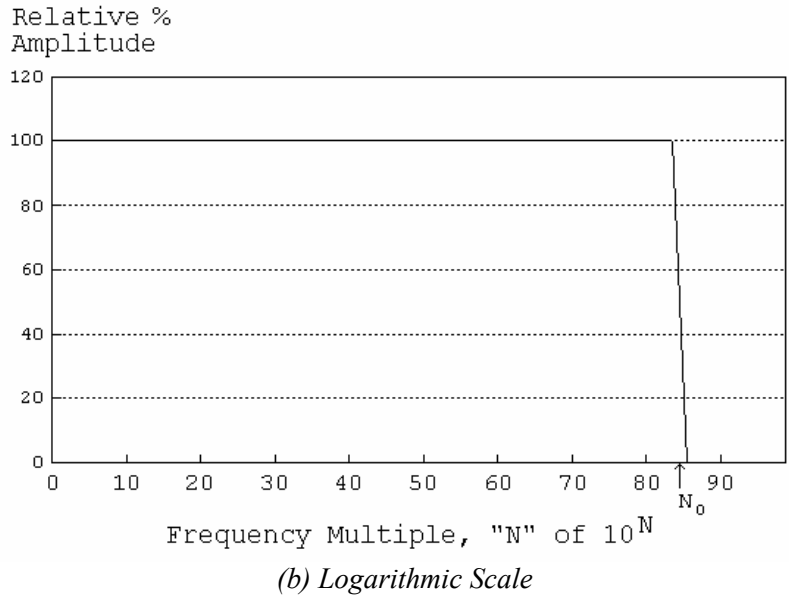

*(b) Logarithmic Scale*
*Figure 10*
*The $Cos^n(x)$ Limitation of the "Cosmic Egg"*

### d *Determination of the Value of $N_0$*

In the original $U(t)$, $N_0$ is the number of protons and electrons (as combined into neutrons) in the original "Cosmic Egg" and that $N_0$, as the exponent of the envelope frequency cosine function, is the effective number of envelopes. The magnitude of that quantity, $N_0$, can be approximately determined. The procedure is to calculate the mass of the universe and divide it by the mass of an individual proton, which is

(28)  $m_p = 1.67\ldots \cdot 10^{-27}$ kilograms.

Hydrogen atoms or their equivalent, that is protons and their associated electrons, are the vast majority, more than *99%* of the matter of the universe. The electron is of negligible mass compared to the proton within the limited accuracy of the present calculation, so it is reasonable here to deem the mass of the universe as being all protons.

Determining the mass of the universe, $m_U$, proceeds by estimating the average mass density, $\rho$, and the applicable universe volume. Those estimates are rather lengthy and involved. The universe mass is the product of the two and its determination by that procedure is developed fully in [4], with the following result for $\rho$.

(29)   $\rho_U \approx 5 \cdot 10^{-27}$ kg/meter$^3$

Next the volume of the universe is needed so as to obtain the universe's mass as the product of the mass density and the volume. The volume of the universe develops as follows. The universe's radius applicable to the just obtained universe mass density should be based on an earlier time than the present because the investigations into estimating that density had to treat astral objects which we observe as they were some time in the past -- their distance from us divided by the speed of their light.

Those earlier times were in the range of *0 to 7 or 8 Gyrs* into the past. As we look into the past at an increasing radial distance from us the observed volumes increase as that radius cubed. For that reason the applicable universe radius to use with the universe mass density just determined is that which existed at the time into the past *t ≈ 6.5 Gyrs* ago. The development in [4] indicates that the estimated radius of the universe for the present calculation is:

(30)   $R_U = 14$ G-Lt-Yrs
             $= 11 \cdot 10^{24}$ meters.

Therefore the mass of the universe, as the product of its volume based on that radius and its equation *(30)* density, is:

(31)   $m_U = \rho_U \cdot [\,^4/_3 \cdot \pi \cdot R_U^3]$
             $= 3 \cdot 10^{49}$ kg.

and the value of $N_0$ from those data is

(32)   $N_0 = \dfrac{m_U}{m_p} = \dfrac{3 \cdot 10^{49}}{1.67 \cdot 10^{-27}} \approx 2 \cdot 10^{76}$

However, analyses in recent years of the hypothesized or speculated likely scenario of the early universe, the "big bang", result in the rough estimate that there were then about $10^9$, one billion, mutual annihilations for every proton present today. (This is based upon the observation that in the present day universe there are about $10^9$ photons per proton. That estimate is a not unreasonable measure of the original number of annihilations. The mutual annihilations each produced two photons. Photons from other later causes, primarily black body radiation and electron orbital changes should be in an amount on the order of one photon per proton, far from $10^9$, and leaving the original mutual annihilations as the dominant source).

In that case the $2 \cdot 10^{76}$ estimate for the present number of particles would give an original $N_0$ value, at the initial instant before any mutual annihilations, of about $2 \cdot 10^{85}$. While all of this estimating is quite approximate it would nevertheless be reasonable to take that $N_0$ was on the order of $10^{85}$.

That is an immense number. And, in this case it is the effective exponent of the envelope *cosine* in *U(t)*; it is the effective number of original envelopes to the "Cosmic Egg". It is the bandwidth limit imposed by the very nature of the original (and on-going) medium's wave oscillation and propagation.

## 5 – *SUMMARY CONCLUSIONS*

1. The universe came into existence from a prior absolute nothing.

   a. The implied infinite rate of change was avoided by the transition from nothing to something being oscillations in a *[1 - cosine]* form.

       b. The implied violation of conservation was avoided by two such oscillations simultaneously beginning, the two identical except of opposite amplitude [+/-].

    2. The complexity of problem 1a, above, led to an apparent infinite set of frequencies in the original oscillation; however, characteristics of the situation limited that set to the finite number $N_0 = 10^{85}$.

       a. One of the limiting factors was a bandwidth limitation.

       b. The other limiting factor was in the nature of the coefficients of successive terms in the expansion of exponentiated cosines.

    3. The expectation that the conservation-maintaining solution should have led to a complete mutual annihilation and no further universe was not fulfilled because of the nature of the complex waveforms.

       a. They represented an immense, complex pair of particles, so unstable that they exploded in an immense radioactive decay into myriad particles, the particles of our universe today. They so exploded before they could completely mutually annihilate.

       b. Our universe, as the evolved successor to that great event, is based on *centers-of-oscillation* of $[1 - cosine]$ form, the protons, electrons, etc. and their anti-particles of our universe.